\documentclass{PoS}

\title{The physics of eight flavours}

\ShortTitle{The physics of eight flavours}

\author{\speaker{Albert Deuzeman} and Elisabetta Pallante \\
         Centre for Theoretical Physics, University of Groningen, 9747 AG Groningen, The Netherlands \\
         E-mail: \email{a.deuzeman@rug.nl}, \email{e.pallante@rug.nl}}
\author{Maria Paola Lombardo \\ 
        INFN-Laboratori Nazionali di Frascati, I-00044, Frascati (RM), Italy \\
        E-mail: \email{mariapaola.lombardo@lnf.infn.it}}

\abstract{When the flavour content of QCD is increased sufficiently, the theory develops a non-trivial infra red fixed point. Thus, for a number of flavours above a certain critical value, but not yet so high that asymptotic freedom is lost, QCD becomes a conformal field theory. The location of the lower limit of this conformal window has not yet been unequivocally determined. Using an improved lattice action, and exploiting modern algorithms allowing for larger lattices and lower quark masses, we have shown that the theory of QCD with eight flavours breaks chiral symmetry in the continuum. We present proof that the accompanying transition is thermal in nature and as a consequence, the conformal window of QCD can only start afterwards, corroborating recent analytical studies at the expense of older results.}

\FullConference{The XXVI International Symposium on Lattice Field Theory \\
                July 14 - 19, 2008\\
                Williamsburg, Virginia, USA}
\PACS{12.38.Gc, 11.15.Ha, 12.38.Mh}

\begin{document}

\section{The conformal window}
In seminal papers, Caswell \cite{Caswell:1974gg} and subsequently Banks and Zaks \cite{Banks:1981nn} showed the existence of an additional zero of the beta function for non-abelian gauge theories with fermionic content in the fundamental representation and for certain values of the number of flavours $N_f$. Their result suggests that QCD like theories with a number of massless flavours within a certain range are both asymptotically free and chirally symmetric at zero temperature. Long range dynamics will therefore exhibit the qualities of a non-trivial conformal field theory. This particular property of non-Abelian field theories is of more than idiosyncratic interest, due to the appearance of conformality in theories beyond the standard model, such as walking technicolour \cite{Miransky:1997, Sannino:2008ha}.

\subsection{Previous results}
The beta function can directly provide us with upper limit of this region, which is the perturbative point where asymptotic freedom is lost. The lower limit depends on non-perturbative dynamics. Several analytic approaches have been applied to this problem, including renormalisation group techniques to estimate the critical point \cite{Appelquist:1998rb} and degree of freedom analysis \cite{Appelquist:1999hr}, arriving at $N_f^c \le 4N\sqrt{1-16/81N^2}$ for an $SU(N)$ theory with $N_f$ flavours. Renormalisation group flow equations predict $N_f^c=10(1)$ \cite{Gies:2005as}. A supersymmetry inspired all-orders beta function leads to a bound of the conformal window \cite{Ryttov:2007cx} of $N_f^c>8.25$ for an SU(3) gauge group. A recent study at finite temperature, on the base of a truncated renormalisation group flow calculation \cite{Braun:2006jd}, found a value of $N_f^c$ close to 12.

On the lattice side, the authors of \cite{Fukugita:1988} performed a pioneering study at $N_f=10$ and $N_f=12$, concluding the former to be outside of the conformal window. Subsequent work by the Columbia collaboration \cite{Brown:1992fz} was not completely conclusive, but tentatively set $N_f=8$ outside of the conformal window. Iwasaki and collaborators, however, found a value as low as $N_f=6$ \cite{Iwasaki:2003de}. In the already mentioned paper \cite{Damgaard:1997ut}, the authors consider $N_f=16$, the largest value of $N_f$ still compatible with asymptotic freedom, and confirm that chiral symmetry is not broken at weak coupling. Several other studies have observed phase transitions as a function of the gauge coupling for $N_f=8$ \cite{Kogut:1982fn, Gavai:1985wi, Kim:1992pk, deForcrand:2007uz}, leaving the question of the fate of these transitions in the continuum limit open. Recent work by Appelquist and collaborators \cite{Appelquist:2007hu}, using the Schr\"odinger functional and a step scaling method, concludes that $N^c_f$ should be close to 12. A new study by Fodor~\textit{et al.} \cite{Fodor:2008hn}, focussing on the spectrum of low eigenvalues for 8 and 12 flavours, concludes cautiously that both are in the regular phase of QCD, allowing for taste breaking effects. Recently, there has also been active interest in the properties of theories using fermions in different representations \cite{Shamir:2008pb, Catterall:2008qk}.

In summary, older studies at $N_f=8$ suggested that the theory is already in the conformal window, while others confirmed analytic calculations \cite{Banks:1981nn, Appelquist:1998rb, Ryttov:2007cx, Braun:2006jd}, indicating that $N_f^c > 8$, close to the upper limit for $N_f^c$ calculated in \cite{Appelquist:1999hr}.

\subsection{Our approach}
Since it is a defining property of the conformal phase, the most straightforward approach to determining if $N_f=8$ is in the conformal window would seem to be measuring the amount of spontaneous chiral symmetry breaking at zero temperature. However, the strong coupling limit of lattice QCD is always confining, regardless of the value of $N_f$ (see again \textit{e.g.}~\cite{Damgaard:1997ut}). So even for those $N_f$ in the conformal window, a phase of broken chiral symmetry exists. However, this strong coupling phase is a lattice artifact and will not be traversed by any physical RG trajectories. When interpreting the simulation results, one therefore has to be certain that the chosen parameters did not put one in an unphysical phase. On the other hand, our finite lattice always introduces some finite temperature. Observing a chirally symmetric vacuum, therefore, could simply mean we are looking at a thermodynamically generated deconfined phase of what is otherwise a confining theory. A single observation of the chiral properties, therefore, does not actually imply anything yet. If a stratagem such as the one proposed here is to work, one needs to determine the nature of the transition itself. What we present here is a determination of the conformal qualities of eight flavour QCD from the nature of the observed phase transition as published previously~\cite{Deuzeman:2008sc}.

\section{Simulation setup}

When $N_f$ is chosen outside of the conformal window, no transition is expected at zero temperature, \textit{i.e.} when the temporal extent of the lattice is effectively infinite. If a lattice transition occurs, it should smoothly move towards $\beta\to\infty$ as $N_t\to\infty$. When it is chosen inside the conformal window, the theory should become insensitive to infrared length scales. Any phase transition at low beta would have to be bulk as a consequence; it should survive the $N_t \to \infty$ limit and be insensitive to its exact value. In practise, one could distinguish between these behaviours by varying the values of $N_t$ and $\beta$ for a given theory. If the critical value of $\beta$ moves appropriately with $N_t$, we are observing a thermodynamical transition at a well defined physical temperature $T_c= \frac{1}{a(\beta_c)\,N_t}$. If it remains stationary, the transition should be bulk. The former option tells us immediately that the theory is not conformal, the latter leaves open the possibility of a transition at weaker coupling.

For our systematic study of the phase transition, we used two values of $N_t$, namely 6 and 12. A third calculation was done at the even smaller value of $N_t=4$, a region that should be problematic for scaling analysis due to the coarse lattice spacing at which the transition should occur there. The value found there was used to generate confidence in our treatment of the $N_t=6$ results. To check for the influence of finite volume effects and deduce the scaling properties of observables, simulations were run for three spatial extents of the lattice $N_s = 12, 20, 24$ for $N_t = 6$, and $N_s=24$ for $N_t=12$. We chose a fixed value of the lattice degenerate quark mass $am = 0.02$ to explore the critical region.

We use an improved Kogut-Susskind fermion action, the ``Asqtad'' action which removes lattice artifacts up to $O(a^2 g^2)$, the details of which can be found in \cite{Deuzeman:2008sc}. The tadpole parameter $u_0$ is the only parameter in need of tuning in this action, and it can determined using its r\^ole in cancelling the the Hartree term in a perturbative expansion~\cite{LM_1985}. As such, it should reflect the expectation value of the physical field strength. Since the improvement feeds back into plaquette value stably, a self consistency criterion suggests itself. This can be implemented rather efficiently on lattices of small sizes and interpolation with low order polynomials produces excellent initial guesses once several values are known. The rational hybrid monte carlo method was employed for the generation of ensembles.

The chiral condensate for $N_f$ degenerate flavours in lattice units
\begin{equation}
a^3\left\langle \bar{\psi} \psi \right\rangle = \frac{N_f}{4N_s^3N_t}\langle \mathrm{Tr} \left[M^{-1}\right] \rangle \, ,
\end{equation}
was determined by using a stochastic estimator with 20 repetitions. The chiral susceptibility, measuring the variation of the chiral condensate with varying the fermion mass $\chi = \partial \langle \bar{\psi} \psi \rangle /\partial m$ at fixed $\beta$ can be divided into a connected and disconnected component $\chi = \chi_{\mathrm{conn}}+\chi_{\mathrm{disc}}$, given in lattice units by
\begin{eqnarray}
a^2\chi_{\mathrm{conn}} &=&  -\frac{N_f}{4 N_s^3 N_t}   \langle \mathrm{Tr} \left[( MM )^{-1}\right ] \rangle                  \nonumber\\
a^2\chi_{\mathrm{disc}} &=& \frac{N_f^2}{16 N_s^3 N_t}\left [  \langle \mathrm{Tr} \left[M^{-1}\right] ^2\rangle - \langle \mathrm{Tr} \left[M^{-1}\right] \rangle^2  \right ]\, ,
\end{eqnarray}
respectively, all written in terms of traces of the staggered fermion matrix $M$. The connected and disconnected contributions to the chiral susceptibility were measured separately, in the same manner as \cite{Bernard:1996zw}.

We can use the chiral susceptibility and the chiral condensate to define two physically relevant quantities
\begin{equation}
\chi_\sigma \equiv\chi = \frac {\partial \langle \bar \psi \psi\rangle}{\partial m} = \chi_\mathrm{conn} + \chi_\mathrm{disc}
\end{equation}
and
\begin{equation}
\chi_\pi = \frac {\langle \bar \psi \psi\rangle}{m}\, .
\end{equation}
They are related through Ward identities to the spacetime volume integral of the
scalar ($\sigma$) and pseudoscalar ($\pi$) propagators and should therefore become degenerate when chiral symmetry is restored. Their associated cumulant $R_\pi \equiv \chi_\sigma/\chi_\pi$ therefore a most useful physical observable in analysing chiral phase transitions.

\section{Results}

\begin{figure}
\center
\includegraphics[width=14 truecm]{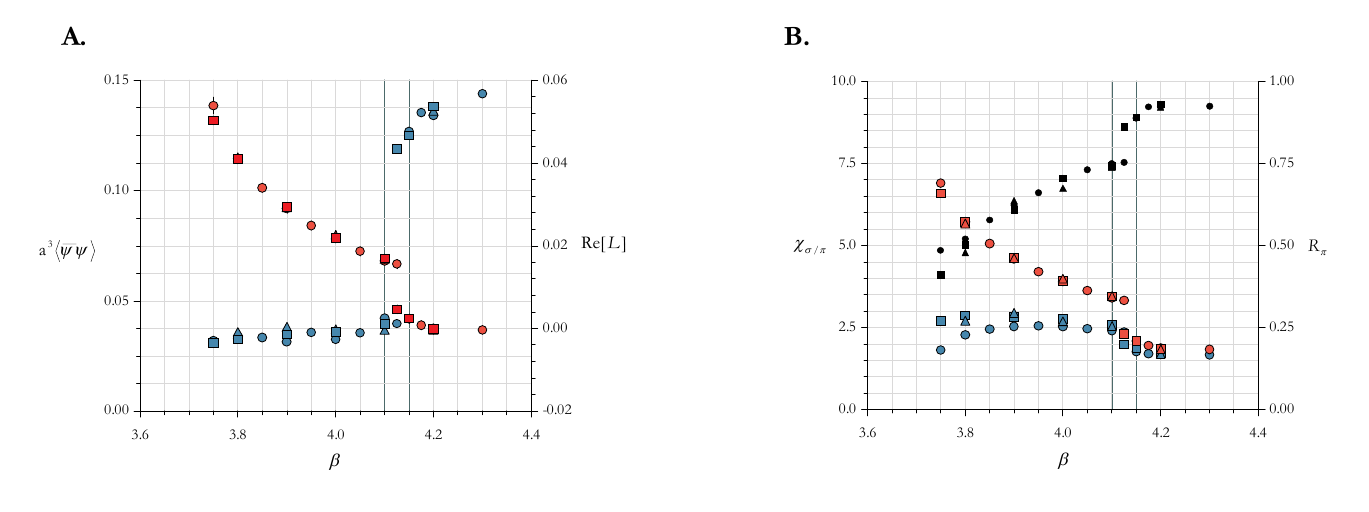}
\caption{\bf A. \rm The chiral condensate (red) and Polyakov loop (blue) as a function of the lattice coupling $\beta$.
\bf B. \rm The scalar $\chi_\sigma$ (blue) and pseudoscalar $\chi_\pi$ (red) chiral susceptibilities as a function of the lattice coupling $\beta$, confirming the degeneracy of the chiral partners in the symmetric phase. The cumulant $R_\pi$ is also shown (black, top of figure).
\bf Both. \rm Measurements were done for three different spatial extents of the lattice $N_s$: 12 ($\bigcirc$), 20 ($\triangle$) and 24 ($\Box$). The critical region has been indicated by vertical lines}
\label{fig:nt_6}
\end{figure}

Results for the chiral condensate (and the Polyakov loop) for $N_t=6$ lattice are shown in figure~\ref{fig:nt_6}A. From the small differences found at different spatial extents, it was concluded that our results can indeed be considered infinite volume estimates for $\beta \le 4.1$ and $\beta \ge 4.15$, and the $\beta$ dependence is smooth. The jump between the two branches is very clear for both observables and suggestive of a discontinuity. A volume dependence is seen in the critical region, as would be expected for a first order phase transition \cite{Karsch:1989pn}. We placed the infinite volume limit value for $\beta_c$ at the lower end of the designated critical region, at and is bounded by $\beta_c = 4.1125(125)$.

The chiral susceptibilities plotted in figure~\ref{fig:nt_6}B confirm our picture. A small amount of splitting remains even in the symmetric phase, which is the explicit breaking induced by the mass term. The cumulant $R_\pi$, as defined earlier and plotted in figure~\ref{fig:nt_6}B shows most clearly the amount of deviation from full chiral symmetry restoration, which lies in the order of 5\%. This observable also exhibits a clearly discontinuous jump within the critical region.

\begin{figure}
\center
\includegraphics[width=8 truecm]{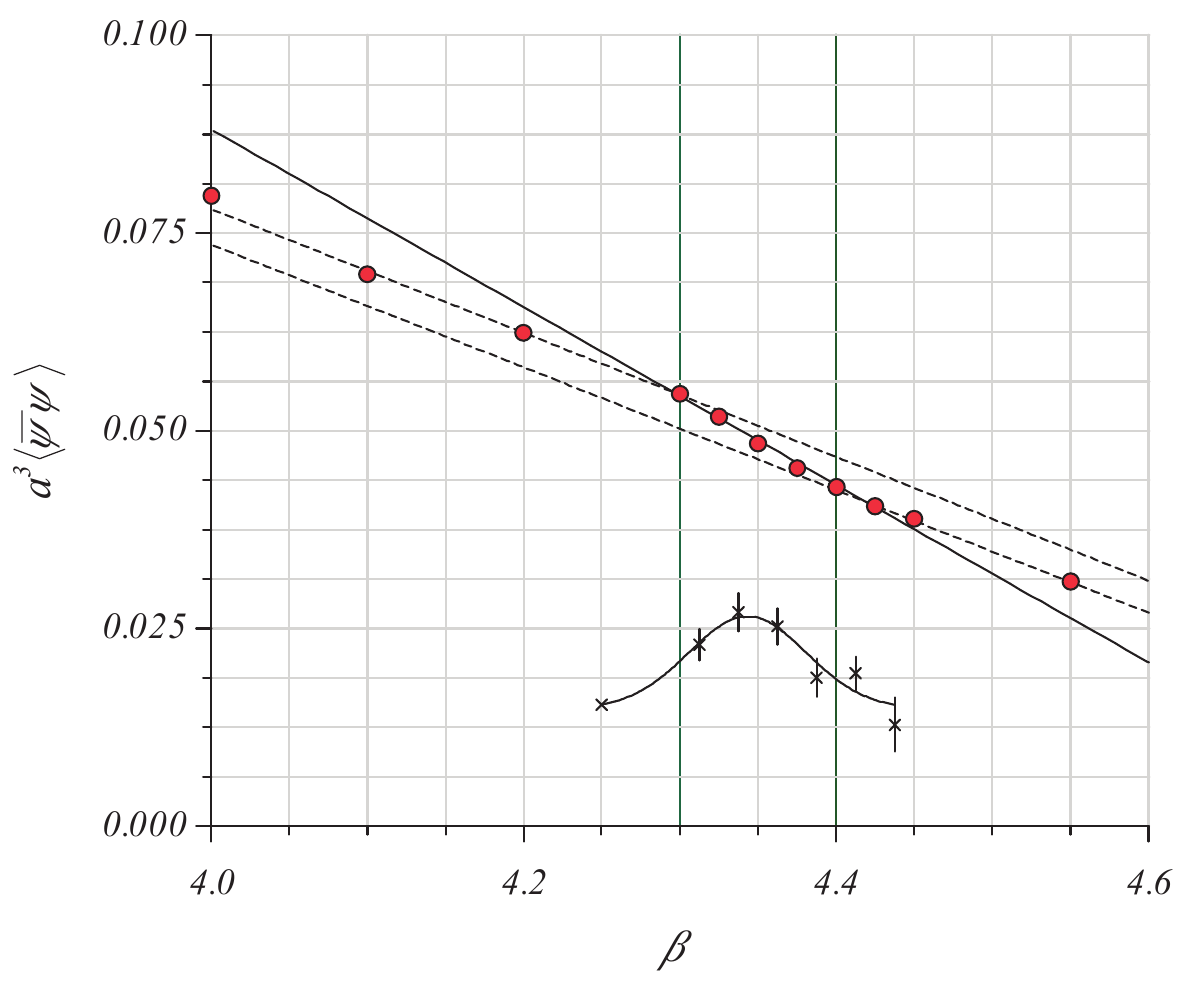}
\caption{The chiral condensate at $N_t=12$ and $N_s=24$ in lattice units  as a function of the lattice coupling $\beta$. Best fit curves are superimposed and the vertical lines indicate the critical region. The absolute value of the finite difference between measured values of the condensate, as an approximation of its first derivative, is plotted in the bottom part of the figure with an arbitrary rescaling. It shows a peak at $\beta = 4.34$.}
\label{fig:pbp_Nt12}
\end{figure}

After having identified the transition point for the coarser lattice, we performed simulations at $N_t = 12$ with spatial volume $N_s=24$ and the same lattice mass $am=0.02$, the main result of which is shown in figure \ref{fig:pbp_Nt12}. We observe a jump in the chiral condensate suggestive of a discontinuity, but slightly distorted and smoothed by the relatively modest spatial extent. Notice also that effects of explicit chiral symmetry breaking due to the non zero physical mass are to be more pronounced here, as the bare fermion mass is now twice as big in physical units. Using a best fit value for the maximum of the derivative indicates an upper bound on the value for the critical coupling of $\beta_c = 4.34(4)$. 

\begin{figure}
\center
\includegraphics[width=14 truecm]{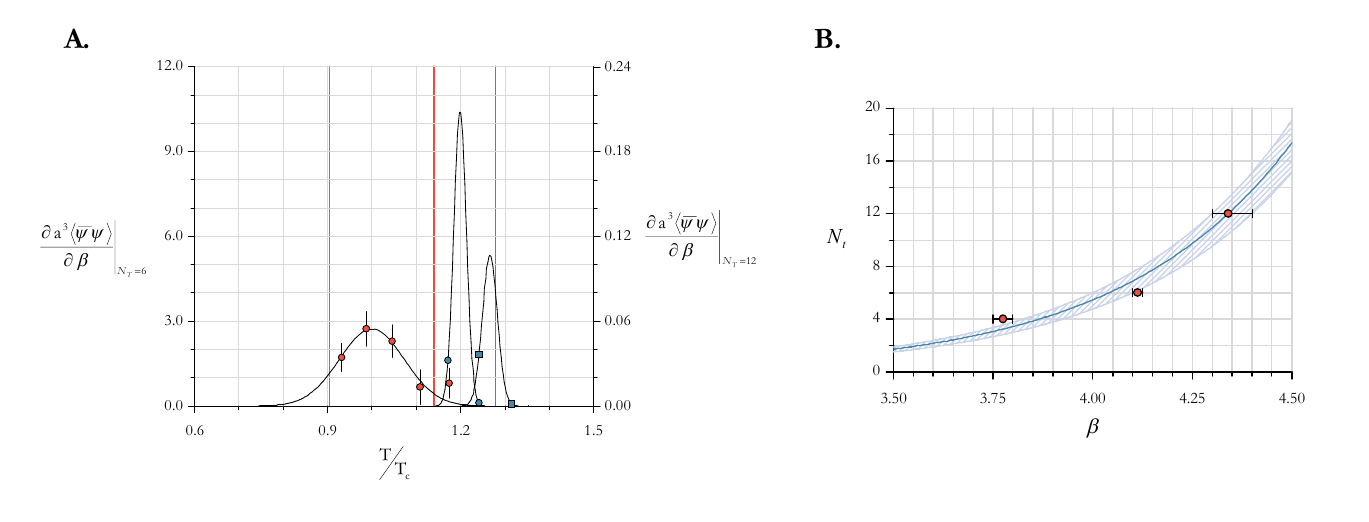}
\caption{\bf A. \rm Finite difference approximation to the absolute value of the first derivative of the chiral condensate as a function of $T/T_c$, determined using perturbative RG scaling. Data at $N_t=6,\, N_s=24, 12$ (blue) are compared with data at $N_t=12,\, N_s=24$ (red). The point $T/T_c=1$ corresponds to $\beta_c(N_t=12) = 4.34$. A Gaussian fit is superimposed to the $N_t=12$ data, while indicative Gaussian curves are shown for $N_t=6$. The baseline is subtracted and we only indicate data near the critical points. The lattice determined critical regions at $N_t=6$ (blue) and $N_t=12$ (red) are indicated by vertical lines.
\bf B. \rm Scaling of the critical temperature presented in terms of the critical lattice coupling $\beta$ for $N_t = 6$ and $12$ lattices. The central line is a two loop asymptotically scaled line of constant physical temperature $T = T_c$, where $N_t = 12$, stemming from the lattice with lowest. All lattice results should eventually follow this line for large enough $N_t$, as lattice artifacts are increasingly suppressed by a decreasing lattice spacing. A preliminary point at $N_t = 4$ has been added for comparison. The lattice spacing here would be expected to be too course for this type of analysis, but the scaling relation seems to hold even up to this point, suggesting effective improvement of the action.}
\label{fig:scaling}
\end{figure}

It is clear that the transition has shifted as a function of the temporal extent of the lattice, which points towards a thermal transition rather than a bulk one. To confirm this, an asymptotic scaling analysis is needed in order to verify that we are actually measuring a critical temperature in the continuum. We do so by means of the standard relation that connects the lattice cutoff $\Lambda_L$ to the gauge coupling $g$, $a\Lambda_L = R(g^2 )$ with
\begin{equation}
\label{eq:2loop}
R(g^2 ) =    (b_0 g^2)^{-{b_1}/{2b_0^2}}\, e^{-1/2b_0 g^2}\, ,
\end{equation}
where the two loop RG running of the $\beta$-function is accounted for, with the universal one- and two-loop coefficients given by
\begin{eqnarray}
b_0 &=& \frac{1}{16\pi^2}\left ( 11-\frac{2}{3} N_f\right ) \nonumber\\
 b_1 &=& \frac{1}{(16\pi^2)^2}\left ( 102-\frac{38}{3} N_f\right )
\end{eqnarray}
for $N_f$ massless flavours. Given the definition $T_c= \frac{1}{a(\beta_c)\,N_t}$, the scaling relation $N_t\, R(g_c (N_t)) = \mathrm{const}$ is implied. Solving for $N_t =6$ and $N_t=12$, we can predict $\beta_c (g_c(N_t=6))$ by knowing $\beta_c (g_c(N_t=12))$. A strong discrepancy with the actual lattice determination might be suggestive of a bulk zero temperature transition, while a small discrepancy can simply be expected and imputed to violations of asymptotic scaling and residual effects due to a non zero fermion mass. We find an extrapolated value of $\beta^{N_t=6}_c = 4.04(4)$, within 2\% of the measured value. In terms of $T/T_c$, this corresponds to good agreement for this sensitive quantity, to within 20\% (see figure \ref{fig:scaling}).
The present results allow us to firmly conclude that we are seeing a true thermal transition, in other words $N_f=8$ undergoes a chiral restoration transition at finite temperature.

\section{Conclusion}

We find $\beta_c(N_f=8,\ N_t=6,\ N_s=\infty) = 4.1125(125)$ and $\beta_c(N_f=8,\ N_t=12,\ N_s = 24) = 4.34(4)$. These values are in agreement with asymptotic scaling according to the perturbative two loop beta function, to about 2\% in terms of $\beta_c$ and to about 20\% in terms of $T_c$. This is fully consistent with the occurrence of a true thermal transition, confirming that $SU(3)$ gauge theory with eight flavours exhibits a chirally broken phase of QCD at zero temperature and in the continuum limit. In agreement with recent analytical results, we conclude therefore that lower bound of the conformal window must lie above $N_f=8$.

\acknowledgments

This work was in part based on the MILC collaboration's public lattice gauge theory code. See http://physics.utah.edu/$\sim$detar/milc.html for details. Computing resources were in part supported by grant nr. SH-079-08 of the Dutch Nationale Computer Faciliteiten (NCF) foundation.

\end{document}